\documentstyle[preprint,aps]{revtex}

\begin{document}
\draft
\title{ Optical responses through dilute anisotropic composites: \\
Numerical calculations via Green's-function formalism }
\author{Y. Gu$^{1,2}$ and K. W. Yu$^2$}
\address{$^1$ State Key Laboratory for Mesoscopic Physics, Department of
Physics, \\
Peking University, Beijing 100871, China }
\address{$^2$ Department of Physics, The Chinese University of Hong Kong, \\
Shatin, New Territories, Hong Kong, China }
\maketitle

\begin{abstract}
We investigate the linear and nonlinear optical responses of dilute
anisotropic networks using the Green's-function formalism (GFF)[Gu Y et al.
1999 {\em Phys. Rev. B} {\bf 59} 12847]. For the different applied fields,
numerical calculations indicate that a large third order nonlinear
enhancement and a broad infrared absorption arise from the geometric
anisotropy. It is also shown the overlap and separation between the
absorption peak and nonlinear enhancement peak when the applied field is
parallel, perpendicular to the anisotropy respectively. In terms of the
inverse participation ratios (IPR) with $q=2$ and spectral distribution of
optical responses, the results can be understood.
\end{abstract}

\pacs{PACC: 7430G, 8160H}

\vskip 5mm

\section{Introduction}

Recently, optical properties of composites have attracted great interest. In
order to open new possibilities in the information processing and
transmission, a large optical nonlinearity may be desirable$^{\cite{bowd71}}$%
. Composite materials, especially, small metal particles embedded in a
dielectric host and metal clusters on the nanometer scale, exhibit a strong
nonlinear optical enhancement through the inhomogeneous local-field and
geometric-response effect$^{\cite{bowd71,yuen41,shal65,siu72}}$. It is also
known that these composites give rise to an anomalously large absorption in
the infrared spectrum$^{\cite{tann25,gran26,deva27}}$. To analyze the
optical responses, it is more convenient to adopt the spectral representation%
$^{\cite{berg3}}$. For various anisotropic composites, the spectral density
was calculated as a function of volume fractions $p_{\Vert }$ and $p_{\bot }$
by the effective-medium approximation (EMA)$^{\cite{yuen41,law42,law43}}$.
It was found a large nonlinear optical enhancement, as well as the
separation of the absorption peak from the nonlinear enhancement peak. By
EMA, the fluctuation of local field has been averaged out. So it is
difficult to find the physical origins of the optical enhancement and the
separation of optical peaks. In this connection, Green's-function formalism
(GFF) was developed to deal with the optical responses of the
arbitrary-shaped metallic clusters embedded in the infinite dielectric
networks at the quasistatic limit$^{\cite{Gu15,thesis16}}$. By the
formalism, the resonance spectrum and local field distribution for each
eigenmode can be analytically obtained. The aim of this paper is to
investigate the linear and nonlinear optical responses throuth the dilute
anisotropic networks in view of the local field distribution.

In the following, a binary dilute anisotropy network is considered. We use
the random generator to produce the geometric anisotropy. The metallic bonds
parallel to the applied field are assigned with the fraction $p_{\Vert }$,
and metallic bonds perpendicular to the applied field with $p_{\bot }$. In
the dilute systems, the condition, $p_{\Vert }\times p_{\bot }=0.01$, is
satisfied. When we change $p_{\Vert }$ from $0.1$ to $0.8$, the
corresponding $p_{\bot }$ varies from $0.1$ to $0.0125$. In Section II, the
Green's functions for the effective linear response $\epsilon _{e}$ and
effective nonlinear response $\chi _{e}$ are derived. In Section III, the
inverse participation ratios (IPR) with $q=2$ are used to represent the
localized and extended eigenstates in anisotropic networks. In Section VI,
the spectra of the absorption and the third order nonlinear enhancement are
illustrated. When the applied field is parallel to the anisotropy of the
networks, the overlap of the absorption peak with nonlinear enhancement peak
is found within the interval $0.1<p_{\Vert }<0.8$. In contrast, for the
perpendicular applied field, the separation of the absorption peak from
nonlinear enhancement peak is enhanced when the anisotropy $p_{\Vert }$ is
increased. These results are explained in Section III, numerically confirmed
in Section VI, and concluded in Section V.

\section{Effective linear and nonlinear responses}

Consider an infinite binary network as shown in Fig. 1, where the impurity
bonds with admittance $\epsilon _{1}$ are employed to replace the bonds in
an otherwise homogeneous network of identical admittance $\epsilon _{2}$.
The admittance of each bond is generally complex and frequency-dependent.
All the impurity bonds construct the clusters subspace. When resonance
happens, the potential of the jointing points can be computed by the GFF.
Instead of the point source $\delta _{{\bf x,0}}^{\cite{Gu15}}$, the source
term is replaced by $\rho _{{\bf x}}$ at the point ${\bf x}$. Hence in the
subspace, $\tilde{V}$ is a linear combination of right eigenvector $\tilde{R}
$'s of Green's-matrix M, 
\begin{equation}
\tilde{V}=\sum_{n=1}^{n_{s}}\frac{s}{\epsilon _{2}(s-s_{n})}\sum_{{\bf y\in C%
}}(\tilde{L}_{n,{\bf y}}\sum_{{\bf x^{\prime }}}\rho _{{\bf x^{\prime }}}%
\tilde{G}_{{\bf y,x^{\prime }}})\tilde{R}_{n}.
\end{equation}%
And for the site ${\bf x}(x_{1},x_{2})$ outside the cluster, $V_{{\bf x}}$
becomes 
\begin{equation}
V_{{\bf x}}=\sum_{{\bf x^{\prime }}}\rho _{{\bf x^{\prime }}}\tilde{G}_{{\bf %
x,x^{\prime }}}+\sum_{n=1}^{n_{s}}\frac{1}{\epsilon _{2}(s-s_{n})}\sum_{{\bf %
y\in C}}(\tilde{L}_{n,{\bf y}}\sum_{{\bf x^{\prime }}}\rho _{{\bf x^{\prime }%
}}\tilde{G}_{{\bf y,x^{\prime }}})\sum_{{\bf z\in C}}M_{{\bf x,z}}\tilde{R}%
_{n,{\bf z}}.
\end{equation}

\begin{figure}[h]
\caption{Schematic diagram of a cluster(shown in thick lines) embedded in an
infinite network}
\end{figure}

In the uniform field $E_{0}$ along ${\bf 1}$ direction, $\tilde{V}$ reads 
\begin{equation}
\tilde{V}=\sum_{n=1}^{n_{s}}\frac{sE_{0}}{\epsilon _{2}(s-s_{n})}(\sum_{{\bf %
y\in C}}\tilde{L}_{n,{\bf y}}y_{1})\tilde{R}_{n}.
\end{equation}%
For a binary network with $N\times N$ square lattices, in the quasistatic
limit, the displacement ${\bf D}$ of the $i$th bond is related to the local
field ${\bf E}$ by the relation ${\bf {D}=\epsilon _{i}{E}+\chi _{i}{|{E}|}%
^{2}{E}}^{\cite{Yu101}}$, where $\epsilon _{i}$ is the dielectric constant
and $\chi _{i}$ is the third order nonlinear susceptibility of the bond. $%
\epsilon _{i},\chi _{i}$ are set to be $\epsilon _{1},\ \chi _{1}$ for the
impurity bonds, and $\epsilon _{2},\ \chi _{2}$ for the matrix bonds
respectively. By the finite difference transformation, the effective linear
response along the applied field is 
\begin{equation}
\epsilon _{e}E_{0}=\sum_{({\bf x,y)\in C}}\frac{\epsilon _{1}(V_{{\bf x}}-V_{%
{\bf y}})}{a}+\sum_{({\bf x^{\prime },y^{\prime })\notin C}}\frac{\epsilon
_{2}(V_{{\bf x^{\prime }}}-V_{{\bf y^{\prime }}})}{a}
\end{equation}%
with the magnitude of field $E_{0}=1$ and lattice constant $a=1$, where $(%
{\bf x,y)\in C}$ means that ${\bf x}$ and ${\bf y}$ are the nearest
neighbors in the same cluster while $({\bf x^{\prime },y^{\prime })\notin C}$
means ${\bf x^{\prime }}$ and ${\bf y^{\prime }}$ are the nearest neighbors
but not in the same cluster. When we employ $N(N+1)\epsilon _{2}=\sum_{{\bf %
x,y}}\epsilon _{2}(V_{{\bf x}}-V_{{\bf y}})$, $\epsilon _{2}-\epsilon _{e}$
can be expressed by the points within the clusters subspace as, 
\begin{equation}
N(N+1)\epsilon _{2}-\epsilon _{e}=\sum_{({\bf x,y)\in C}}(\epsilon
_{2}-\epsilon _{1})(V_{{\bf x}}-V_{{\bf y}}).
\end{equation}%
Here $\epsilon _{1}/\epsilon _{2}=(s-1)/s$, so we have 
\begin{equation}
N(N+1)-\frac{\epsilon _{e}}{\epsilon _{2}}=\sum_{n=1}^{n_{s}}\frac{1}{%
\epsilon _{2}(s-s_{n})}(\sum_{{\bf y\in C}}\tilde{L}_{n,{\bf y}}y_{1})\sum_{(%
{\bf x,y)\in C}}(x_{1}-y_{1})(\tilde{R}_{n,{\bf x}}-\tilde{R}_{n,{\bf y}}).
\end{equation}%
Because ${\bf x}$ and ${\bf y}$ are neighbors, the values of $x_{1}-y_{1}$
can be only $+1$ or $-1$. Physically, the absorption along the applied field
is larger than that along the perpendicular direction of the applied field.
In the following calculations, $\epsilon _{e}$ always represents the
absorption along the applied field. For simplicity, we let $L_{\hat{n}%
}=\sum_{{\bf y\in C}}\tilde{L}_{n,{\bf y}}y_{1}$ , $R_{\hat{n}}=\sum_{{\bf %
x,y\in C}}(x_{1}-y_{1})(\tilde{R}_{n,{\bf x}}-\tilde{R}_{n,{\bf y}}),\ $ and 
$\gamma _{n}={\it L_{\hat{n}}R_{\hat{n}}}$, the imaginary part of $\epsilon
_{e}$ corresponding to the absorption is 
\begin{equation}
-Im(\epsilon _{e})=Im\sum_{n=1}^{n_{s}}\frac{\gamma _{n}}{\epsilon
_{2}(s-s_{n})}.
\end{equation}%
$\gamma _{n}$, referred to as the cross section, obeys the sum rule$^{\cite%
{clerc55}}$, 
\begin{equation}
\sum_{n}\gamma _{n}=N_{h}
\end{equation}%
where $N_{h}$ is the number of horizontal bonds along the applied field.

Since the local fields are determined completely, our formulas can be used
to calculate the nonlinear response. By relating the total electrostatic
energy to the effective coefficients, the third order nonlinearity $\chi
_{e}^{\prime }$ is defined by$^{\cite{Yu101}}$%
\begin{equation}
\int_{v}{\bf D(x)}\dot{{\bf E(x)}}=V[\epsilon _{e}^{\prime }{\bar{{\bf E}}}%
^{2}+\chi _{e}^{\prime }{\bar{{\bf E}}}^{4}],
\end{equation}%
where $\bar{{\bf E}}=(1/V)\int_{v}{\bf E(x)d^{3}x}$ is the space averaged
electric field. For the infinite networks, $\bar{{\bf E}}={\bf E_{0}=1}$ and
the nonlinear response function is given by$^{\cite{law42}}$. 
\begin{equation}
\chi _{e}^{\prime }=\frac{1}{l^{2}}\sum_{i}\chi _{i}|\delta v_{i}|^{2}{%
\delta v_{i}}^{2},
\end{equation}%
where the summation is over all bonds and $\delta v_{i}$ is the(general
complex) potential difference across the bond $i$ and for two dimensional
case, $V=l^{2}$. When all the bonds have nonlinear term, i.e., $\chi
_{1}=\chi _{2}=1.0$, with the lattice constant $a=1$, the effective
nonlinear response is written as 
\begin{equation}
\chi _{e}=V\chi _{e}^{\prime }=\sum_{({\bf x,y)}}|V_{{\bf x}}-V_{{\bf y}%
}|^{2}(V_{{\bf x}}-V_{{\bf y}})^{2}.
\end{equation}%
The third order nonlinearity $\chi _{e}$ is expressed as the summation of
the forth moment of local electric field. So the fluctuation of local
electric field enhances the nonlinear optical responses well.

\section{Inverse participation ratios with $q=2$}

First, the IPR of eigenvectors of Green's-matrix $M$ in the eigensystem with 
$n_{s}$ jointing points are defined. The $n$th normalized right eigenvector
is 
\begin{equation}
R_{n}=\{{R_{n,1},\ .\ .\ .\ ,\ R_{n,i},\ .\ .\ .\ R_{n,n_{s}}\}}
\end{equation}%
with $<R_{n}>=0$ and $<{R_{n}}^{2}>=1$. The IPR of $R_{n}$ is written as$^{%
\cite{Wegner}}$, 
\begin{equation}
\mbox{IPR}(R_{n})=\sum_{i=1}^{n_{s}}{R_{n,i}}^{2q}.
\end{equation}%
Here $q=2$. The calculation of IPR$(R_{n})$ is limited in the nontrivial
eigenstates. The number of IPR$(R_{n})$ is equal to or less than $n_{s}$. In
the above Eqs. (1) and (2), it is found that the right eigenvectors of $M$
are closely related to the local fields of the impurity cluster in the
subspace. So the IPR can be used to represent the localization of the
eigenstates. The IPR amplify the profiles of eigenstates, namely, the
localized states become more pronounced and extended states become smoother.
Hence, the larger values of IPR are always corresponding to the stronger
optical responses. This will be verified in the following sections.

To consider the size effect, the IPR in the dilute isotropic composites with 
$p_{\Vert }=0.1$ and $p_{\bot }=0.1$ are shown in Fig. 2. In this figure,
two samples are in size $30\times 30$ and $60\times 60$. The peaks represent
the localized states and the valleys correspond to the extended states.
Comparing the distribution of peaks and valleys in two samples, we find that
the size effect is not very obvious. It is also seen that the density of
states are larger around $s=0.5$ than that around $s=0.0$ or $s=1.0$. The
localized states incline to accumulate $s=0.0$ or $s=1.0$, while, the
extended states are near $s=0.5$. So there exists a duality about $s=0.5^{%
\cite{thesis16}}$. The high values of IPR imply the strong optical responses
at $s=0.0$ or $s=1.0$.

\[
\]%
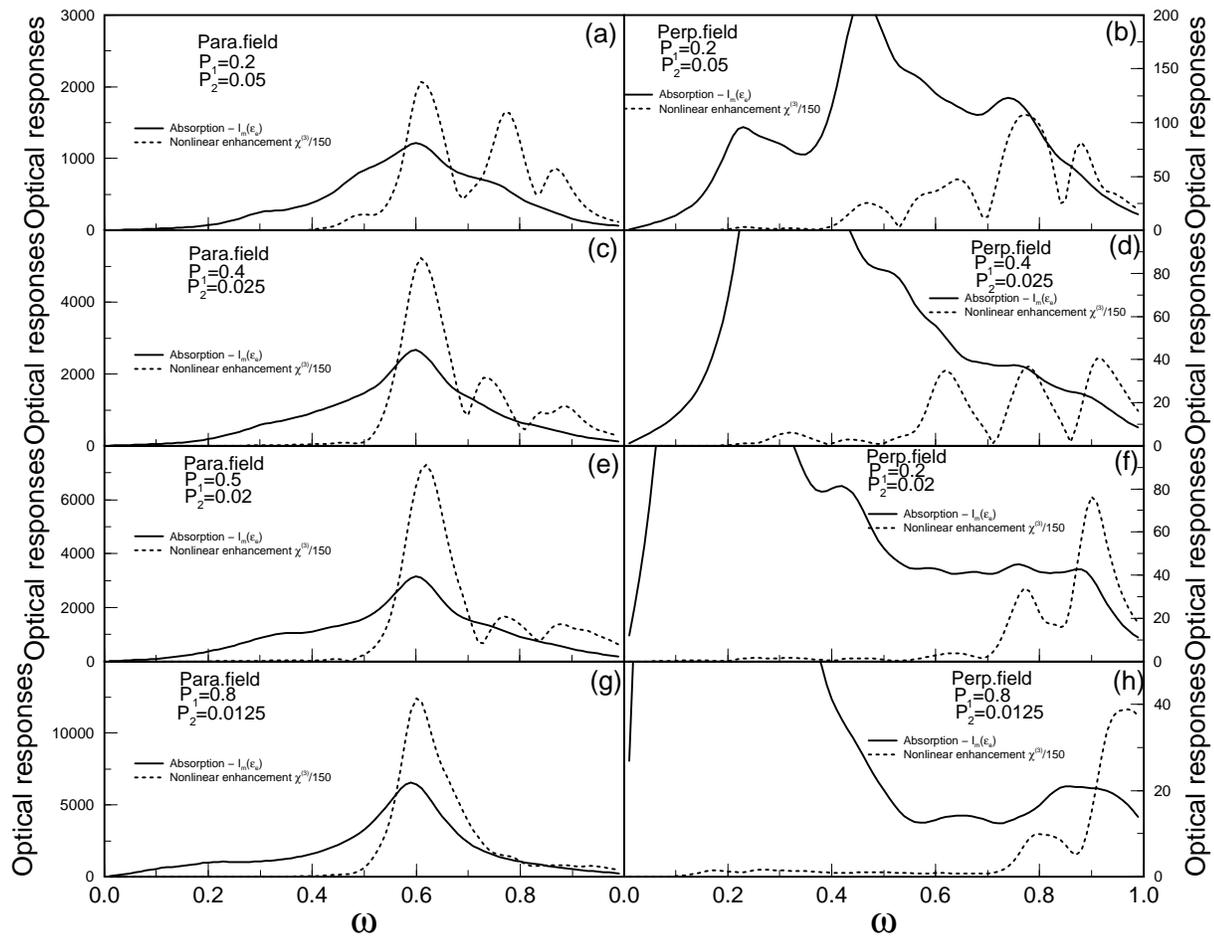
\begin{figure}[h]
\caption{ Size effect of IPR of right eigenvectors for two samples: $%
30\times 30$ and $60\times 60$. Here $p_{\Vert }=p_{1}=0.\ 1,\ p_{\bot
}=p_{2}=0.\ 1$. }
\end{figure}

Fig. 3 displays the IPR of the dilute anisotropic systems. We have known
that the resonance spectrum is very sensitive to the microstructure$^{\cite%
{thesis16}}$. For each case, comparing two samples, we see that the peaks of
IPR are very stable though their microstructure is completely different. So
in the following, for the definite parameters $p_{\Vert }$ and $p_{\bot }$,
it is reasonable to investigate the optical responses by only one sample.
When the anisotropy is increased from Fig. 3(a) to Fig. 3(g), the localized
states incline towards $s=0.\ 0$ or $s=1.\ 0$. The distributions of
localized and extended states have roughly the same feature as that in the
isotropic case, i.e., we find more localized states at $s=0.0$ and $s=1.0$
than at $s=0.5$, but denser at $s=0.5$. The distribution of IPR is caused
purely by the morphology of the sample. In the next section, we will discuss
how the different applied fields act on the optical responses when the
geometric anisotropy is increased.

\[
\]

\begin{figure}[h]
\caption{ IPR of right eigenvectors of the dilute anisotropic systems.
Sample 1 and sample 2 are randomly chosen. In (a) and (b), $p_{\Vert
}=p_1=0.\ 2,\ p_{\bot }=p_2=0.\ 05$. In (c) and (d), $p_{\Vert }=p_1=0.\ 4,\
p_{\bot }=p_2=0.\ 025$. In (e) and (f), $p_{\Vert }=p_1=0.\ 5,\ p_{\bot
}=p_2=0.\ 02$. In (g) and (h), $p_{\Vert }=p_1=0.\ 8,\ p_{\bot }=p_2=0.\
0125 $. }
\end{figure}

\section{Spectra of absorption and third order nonlinear enhancement}

The optical responses can be properly described by the absorption spectrum $%
\gamma _{n}$ and the third order nonlinear enhancement spectrum as shown in
Fig. 4 and Fig. 5. When the applied field is parallel to the direction of
anisotropy, it affects the optical properties well. The envelopes of the
absorption and third order nonlinear enhancements become narrower and the
intensity is more enhanced with the increasing anisotropy. The main optical
peaks are ranged in the interval $s\in \lbrack 0.5,0.6]$, which correspond
to the dipolar approximation. In this figure, we can not find the separation
of the absorption peak from the third order nonlinear enhancement peak. The
results of single sample conflict with those of previous works$^{\cite%
{thesis16}}$. The reason is that the EMA has averaged out the fluctuation of
local field, which plays the central role for the effective optical
responses. At $s=0.5$, the density of localized states are much smaller than
that of extended states. The optical properties of the specific sample are
determined by its own eigenstates, not by ``effective medium''. So it is not
reasonable for all of the cases to express the optical properties of one
specific sample only by the same macroscopic parameters $p_{\Vert }$ and $%
p_{\bot }$. In Fig. 5, when the applied field is perpendicular to the
anisotropy, the optical responses of the dilute anisotropic samples are
plotted. It is seen that the peaks of absorption are red-shifted while the
peaks of the third order nonlinear enhancement are blue-shifted when the
anisotropy is added. There is a larger peak separation with increasing
geometric anisotropy. For all the applied fields, it is natural to find a
large third order nonlinear enhancement and a broad infared absorption
because the localized states exist in the whole resonant area. Only for the
perpendicular applied field, the separation of optical peaks is found.

\[
\]

\begin{figure}[h]
\caption{ Spectra of the absorption and the third order nonlinear
enhancement of the dilute anisotropic systems for the parallel applied
field. In (a) and (b), $p_{\Vert }=p_{1}=0.\ 2,\ p_{\bot }=p_{2}=0.\ 05$. In
(c) and (d), $p_{\Vert }=p_{1}=0.\ 4,\ p_{\bot }=p_{2}=0.\ 025$. In (e) and
(f), $p_{\Vert }=p_{1}=0.\ 5,\ p_{\bot }=p_{2}=0.\ 02$. In (g) and (h), $%
p_{\Vert }=p_{1}=0.\ 8,\ p_{\bot }=p_{2}=0.\ 0125$. }
\end{figure}

\[
\]

\begin{figure}[h]
\caption{ Spectra of the absorption and the third order nonlinear
enhancement of the dilute anisotropic systems for the perpendicular applied
field. In (a) and (b), $p_{\Vert }=p_1=0.\ 2,\ p_{\bot }=p_2=0.\ 05$. In (c)
and (d), $p_{\Vert }=p_1=0.\ 4,\ p_{\bot }=p_2=0.\ 025$. In (e) and (f), $%
p_{\Vert }=p_1=0.\ 5,\ p_{\bot }=p_2=0.\ 02$. In (g) and (h), $p_{\Vert
}=p_1=0.\ 8,\ p_{\bot }=p_2=0.\ 0125$. }
\end{figure}

Then, the Drude free electronic model is employed to calculate the optical
properties of the dilute anisotropic composites. The admittance of the
impurity metallic bonds is 
\begin{equation}
\epsilon _{1}=1-\frac{\omega _{p}^{2}}{\omega (\omega +i\gamma )},
\end{equation}%
where $\omega $ is the plasma frequency, and $\gamma $ a damping constant.
For metal, the plasma frequency $\omega _{p}~10^{16}$, being in the
ultraviolet. We choose $\gamma =0.1\omega _{p}$, which is the typical value
of a metal, and $\epsilon _{2}=1.77$, which is the dielectric constant of
water for model calculations. The range of optical responses is $\omega
/\omega _{p}\in (0,1)$.

Optical responses of various anisotropic networks are illustrated in Fig. 6.
The nonlinear enhancement $\chi ^{(3)}$(or $\chi _{e}$) is divided by $100$
for the comparison with the linear absorption. The direction of anisotropy
in Figs. 6(a), (c), (e) and (g) is parallel to the applied field. The peak
at $\omega /\omega _{p}=0.\ 6$ corresponds to the dipolar approximation of
the whole system. In these figures, the separation of absorption peak and
nonlinear enhancement peak is not found, as observed in the above Fig. 4.
Note that this result is different from that of the previous works$^{\cite%
{yuen41,law42,law43}}$. However, when the applied field is perpendicular to
the anisotropy, as shown in Fig. 6(b), (d), (f) and (h), in each of which,
both the separation of linear and nonlinear peaks, and the third order
nonlinear enhancement are seen.

\[
\]

\begin{figure}[h]
\caption{ Linear and nonlinear optical responses of the dilute anisotropic
systems by Drude model. Para.field means that the applied field is parallel
to the anisotropy while perp.field is that the applied field is
perpendicular to the anisotropy. Here $p_1$ is corresponding to $p_{\Vert }$
in the text and $p_2$ is $p_{\bot }$. }
\end{figure}

\section{Discussion and conclusions}

In this paper, optical responses of dilute binary anisotropic networks are
investigated numerically by the GFF. Green's functions for the effective
linear and nonlinear responses are derived. The IPR with $q=2$ are given to
describe the localization and extension of eigenstates. The spectra of the
absorption and the third order nonlinear enhancement are exhibited for the
different applied fields. We also compute the optical responses for the
Drude model. For the dilute anisotropic networks, we conclude that:

\begin{enumerate}
\item The size effect of optical responses is not obvious.

\item The IPR with $q=2$ can be used to represent the localization and
extension of eigenstates.

\item The wide optical absorption and large nonlinear enhancement are caused
by the geometric anisotropy.

\item The peaks of the absorption and nonlinear enhancement overlap when the
applied field is parallel to the anisotropy. While, for the perpendicular
applied field, the absorption peak is separated from the nonlinear
enhancement peak.

\item The structure sensitivity does not affect the main optical properties,
but the details of optical responses.
\end{enumerate}

>From above 2, 3, 4 and the previous works$^{\cite{yuen41,law42,law43}}$,
geometric anisotropy is the main reason to enhance the optical responses.
For a specific sample, when the different fields are applied, the positions
of optical responses do not vary, but the local field distributions are
quite different. Therefore the optical properties of resonant composites are
determined by the geometric anisotropy, as well as by the applied sources.

\section*{Acknowledgements}

Gu Y acknowledges the useful discussions with Prof. Yang Zhan Ru.

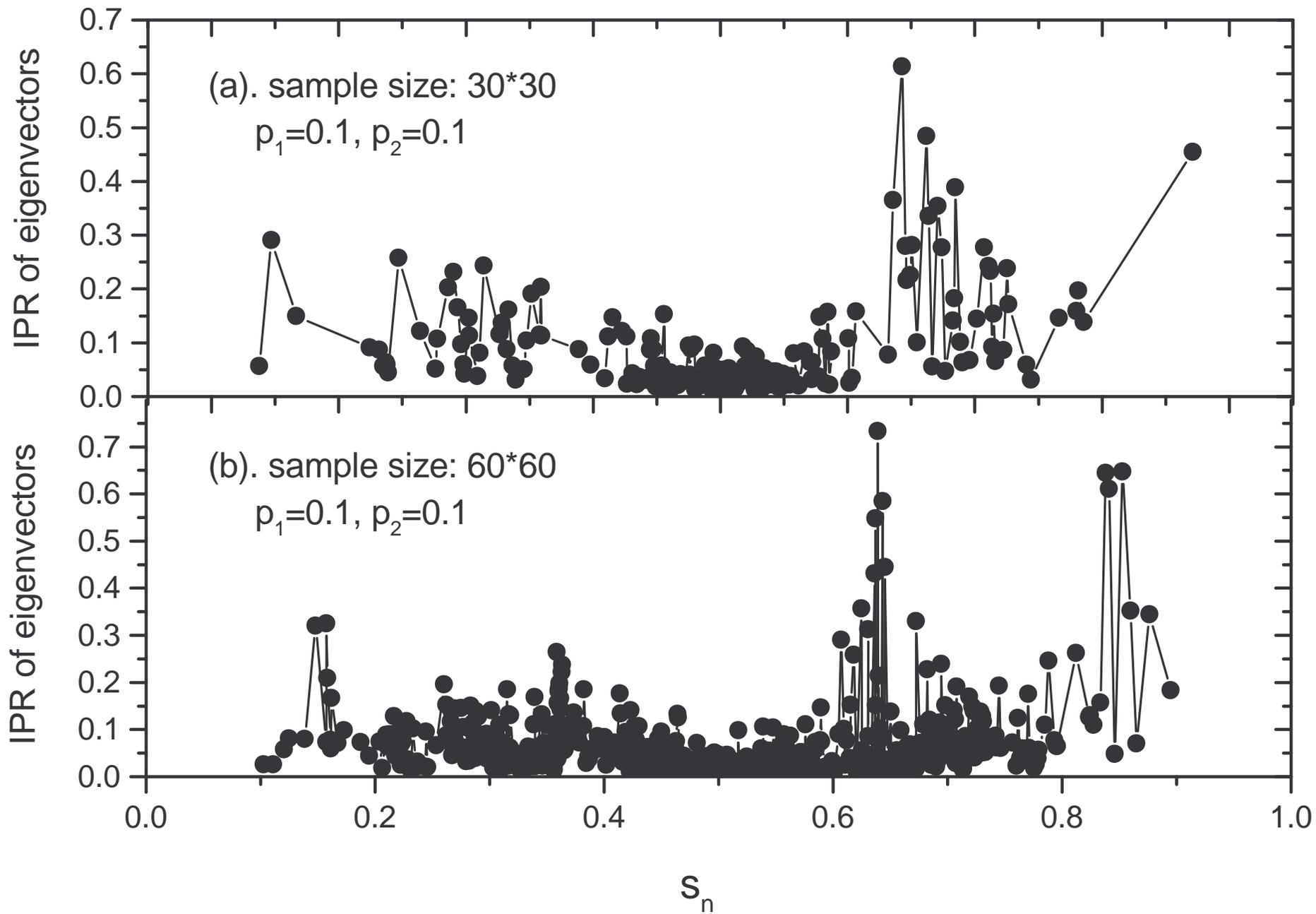
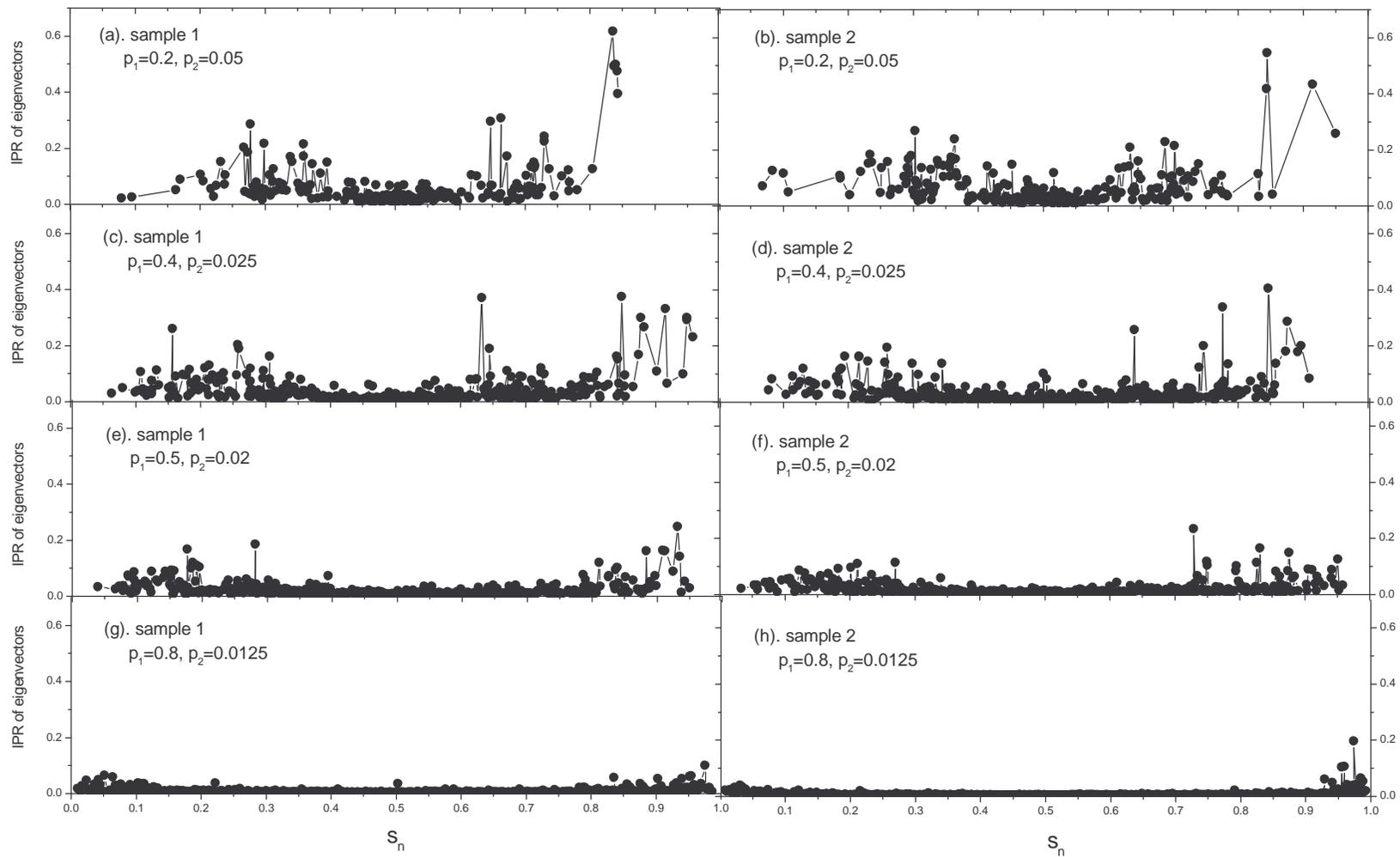
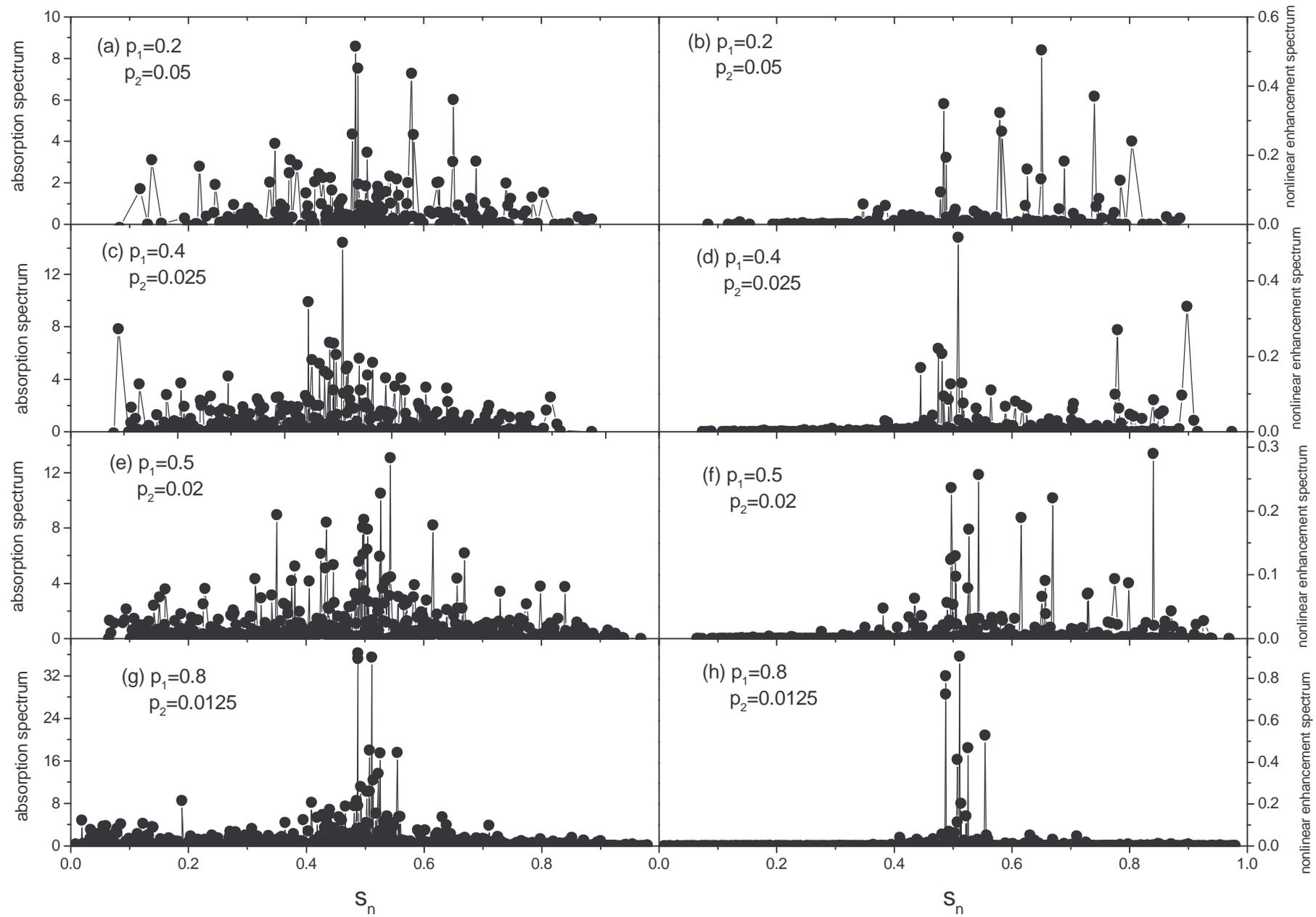
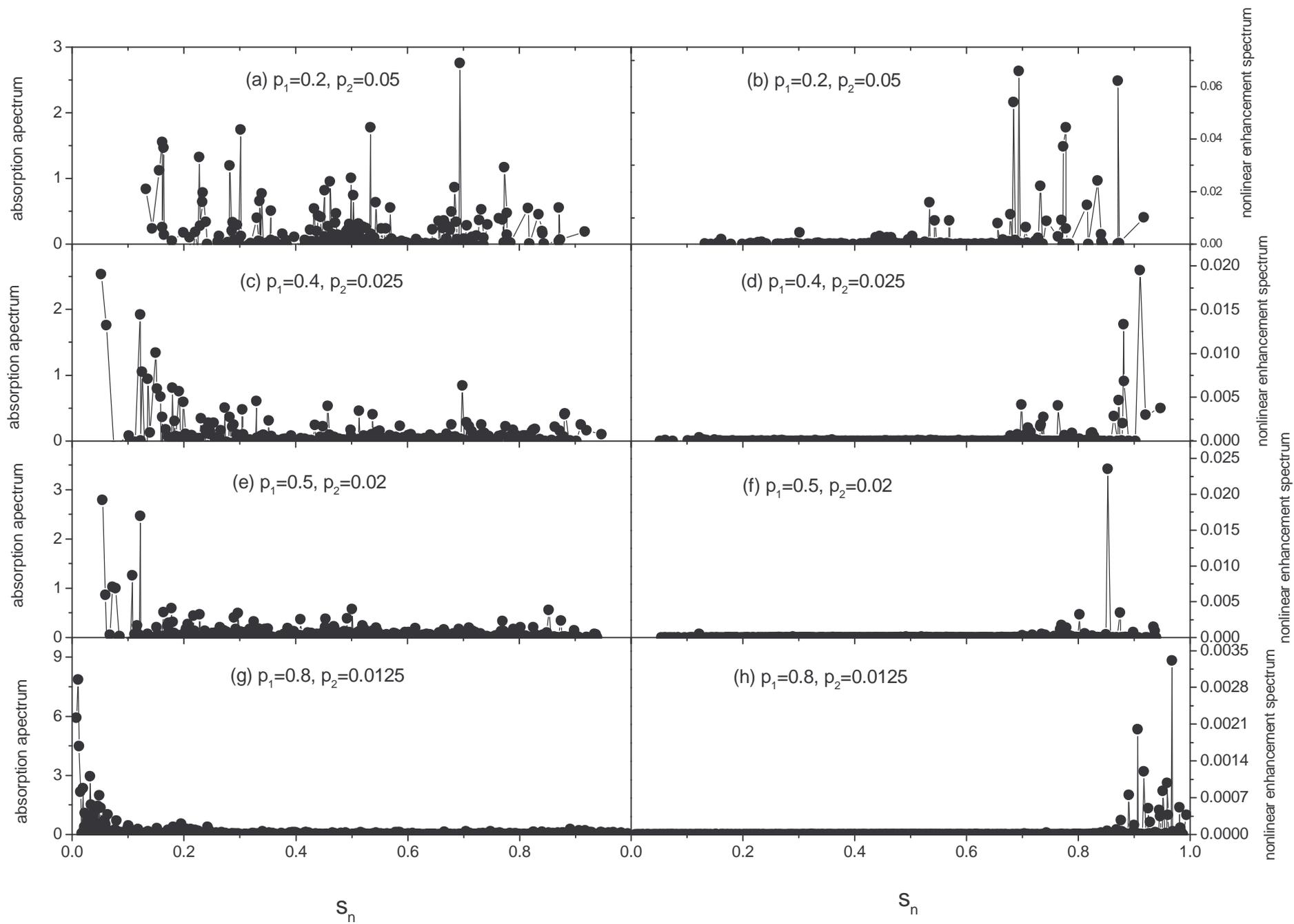
\end{document}